\documentclass[aps,prl,twocolumn]{revtex4-2}
\usepackage{amssymb}
\usepackage{graphicx,amsmath}
\usepackage{hyperref}
\usepackage{color}
\allowdisplaybreaks
\newcommand{\bea}{\begin{eqnarray}}
\newcommand{\eea}{\end{eqnarray}}
\newcommand{\beq}{\begin{equation}}
\newcommand{\eeq}{\end{equation}}
\newcommand{\benu}{\begin{enumerate}}
\newcommand{\enu}{\end{enumerate}}
\newcommand{\la}{\langle}
\newcommand{\ra}{\rangle}
\newcommand{\al}{\alpha}
\newcommand{\be}{\beta}
\newcommand{\ga}{\gamma}

\newcommand{\om}{\omega}
\newcommand{\Om}{\Omega}
\newcommand{\ep}{\epsilon}

\newcommand{\dl}{\delta}

\newcommand{\tht}{\theta}
\newcommand{\ham}{\mathcal{H}}

\newcommand{\prm}{\prime}
\newcommand{\ptl}{\partial}

\newcommand{\cda}{c^{\dagger}}
\newcommand{\bk}{{\bf k}}
\newcommand{\bq}{{\bf q}}

\newcommand{\bQ}{{\bf Q}}
\newcommand{\hGa}{\hat{\Gamma}}
\newcommand{\hK}{\hat{K}}
\newcommand{\hH}{\hat{H}}
\newcommand{\hv}{\hat{v}}
\newcommand{\hx}{\hat{x}}
\newcommand{\hy}{\hat{y}}
\newcommand{\hh}{\hat{h}}

\begin{document}

\title{
Antisymmetric Raman response}
\date{\today}
\author{Mattia Udina$^{1,2}$ and Indranil Paul$^1$}
\affiliation{
$^1$Laboratoire Mat\'{e}riaux et Ph\'{e}nom\`{e}nes Quantiques,
Universit\'{e} Paris Cit\'{e}, CNRS, 75205 Paris, France\\
$^2$Institut de Physique et Chimie des Mat\'{e}riaux de Strasbourg (UMR 7504),
Universit\'{e} de Strasbourg and CNRS, Strasbourg, 67200, France
}

\begin{abstract}
We develop the theory of antisymmetric Raman response, defined as
the difference between the Raman signals of two scattering
geometries related by an exchange of mutually perpendicular incoming
and the outgoing photon polarizations. Such responses, finite in
orthorhombic or lower symmetry systems, are related to
cross-susceptibilities of two Raman operators and are characterized
by the absence of intraband terms. This is in contrast to standard
Raman responses which measure auto-susceptibilities where both
intra- and interband processes contribute. We illustrate the theory
with examples from the charge density wave rare-earth tritellurides
and the excitonic insulator Ta$_2$NiSe$_5$. Our theory establishes
antisymmetric Raman response as a unique tool to probe microscopic
features such as interband energy scales and to detect reflection
symmetry breaking.
\end{abstract}

\maketitle

\emph{Introduction.}---
Inelastic light scattering, with distinct incoming and outgoing photons, or Raman response is an important
tool to probe quantum
materials~\cite{Abrikosov61, Abrikosov74, Klein82, Balkanski83, Klein84,
Shastry90, Shastry91, Hayes04, Devereaux07,Sacuto13,Gallais16}.
The versatility of this technique
stems from the ability to control the photon polarizations.
In conventional usage, Raman response
measures the auto-susceptibility of a second rank tensor operator $\hGa$, whose precise tensorial structure
is determined by the aforesaid polarizations.
\begin{figure}[b]
\begin{center}
\includegraphics[width=0.45\textwidth]{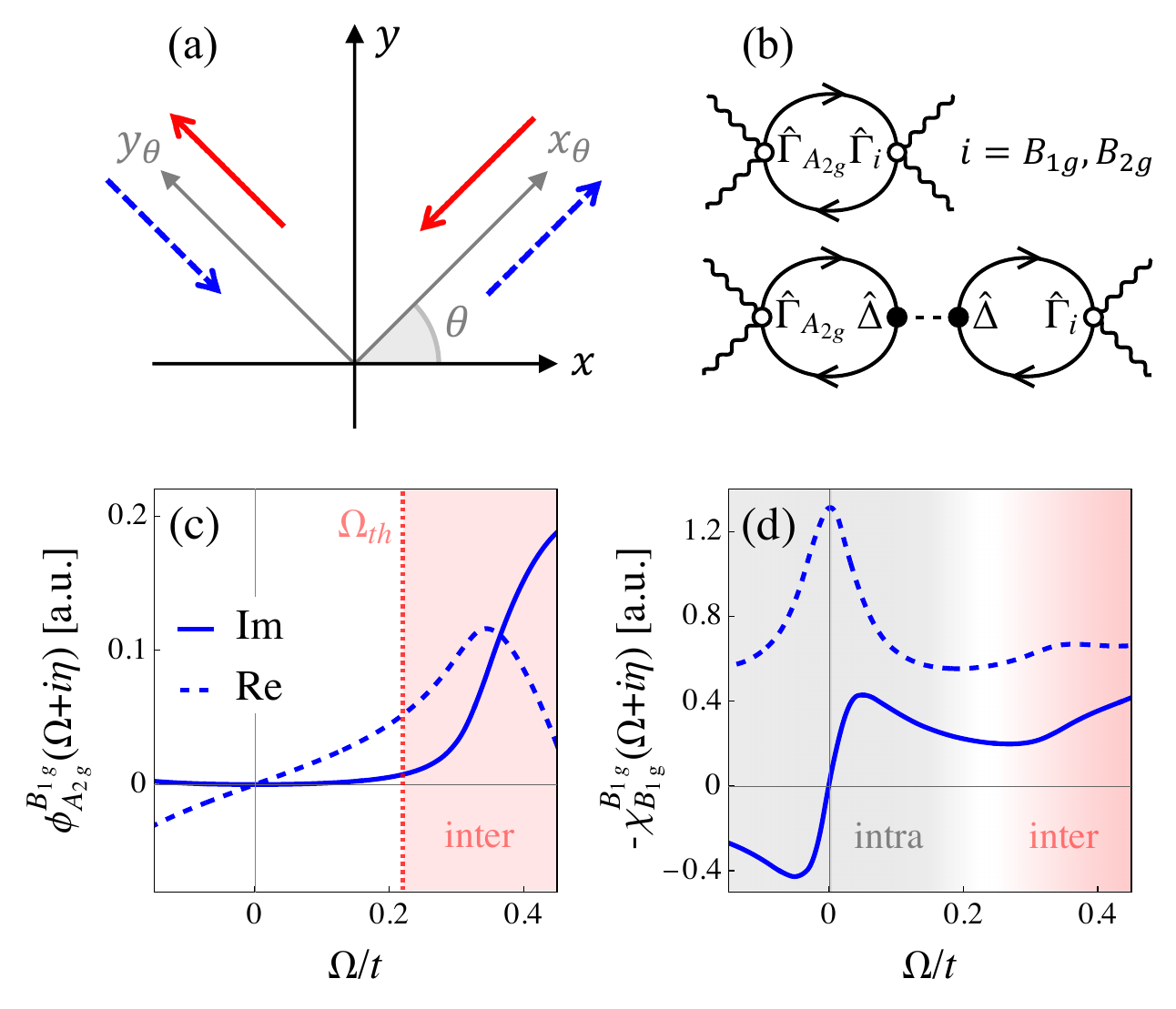}
\caption{(Color Online) (a) Raman
intensity $ I_{x_{\tht}, y_{\tht}}(\Om)$ with mutually perpendicular in, out photon
polarizations (red arrows). $ I_{y_{\tht}, x_{\tht}}(\Om)$ has polarizations exchanged (blue arrows). (b) Graphs of
antisymmetric Raman responses from electrons (one loop) and a collective mode (two loop).
$\hGa$ are Raman operators (white bullets), $\hat{\Delta}$ is electron (solid line) - collective mode (dashed line)
coupling (black bullets). Comparing $\phi_{A_{2g}}^{B_{1g}}$ in (c) with
$\chi_{B_{1g}}^{B_{1g}}$ in (d) for an orthorhombic metal. In (c) the imaginary and real parts are even and
odd in $\Om$, respectively. In (d) it is opposite. Grey, red shades are intra- and interband dominated,
respectively. ${\rm Im} \phi_{A_{2g}}^{B_{1g}}(\Om)$, lacking intraband component,
is zero below threshold $\Om_{th}$. }
\label{fig1}
\end{center}
\end{figure}

In this work we describe a new application of Raman scattering that probes the \emph{cross-susceptibility}
of two Raman operators $\hGa_1$  and $\hGa_2$. It is realized by comparing the signal
$I_{x_{\tht}, y_{\tht}}$ with mutually perpendicular
in, out polarizations $(x_{\tht}, y_{\tht})$, respectively, with $ I_{y_{\tht}, x_{\tht}}$ having the polarizations
exchanged, see Fig.~\ref{fig1}(a). We define
$I_A (\Om, \tht) \equiv I_{x_{\tht}, y_{\tht}}(\Om) - I_{y_{\tht}, x_{\tht}}(\Om)$ as \emph{antisymmetric Raman response}.
They are non-zero in orthorhombic or lower symmetry systems, as discussed below.

Our motivation is a recent experiment on the rare-earth tritellurides~\cite{Burch22}, where a
noticeable $I_A (\Om)$, $\Om \sim $ meV,
due to a collective mode was reported in the charge density wave (CDW) phase.
In Ref.~\onlinecite{Burch22} it was attributed to eV scale resonant electron scattering,
which is contrary to separation of energy scales argument.
Therefore, our first goal is to show that this signal is present in $I_A (\Om, \tht)$
by considering only low-energy \emph{nonresonant}
scattering, provided \emph{all} reflection symmetries  are broken. 
This crucial symmetry breaking ingredient is missing in the theory of Ref.~\onlinecite{Burch22}.
Our second goal is to study the general properties of $I_A (\Om, \tht)$, show them to be quite
different from conventional Raman signals, and to establish $I_A (\Om, \tht)$ 
as a new tool to probe electronic systems.

The following are our main results.
(i) We connect $I_A (\Om, \tht)$ to finite frequency cross-susceptibilities of two Raman operators
by a fluctuation-dissipation relation.
(ii) We show that the cross-susceptibilities have spectroscopic ``gaps'' due to complete absence of 
intraband terms even in metals.
(iii) We show that the frequency integrated cross-susceptibilities are thermodynamic
order parameters for reflection symmetry breaking. (iv) We obtain a reciprocity relation that
$I_A (\Om, \tht)$ is an even function of an external magnetic field. (v) Using examples motivated by the
tritellurides~\cite{Burch22,Burch24,Wulferding24,Lavagnini10,Brouet04,Brouet08,Moore10,Bolloch24}
and the exciton insulator Ta$_2$NiSe$_5$~\cite{Salvo86,Lu17,Mazza20,Ohta25},
we show that antisymmetric Raman response contain information that
cannot be accessed using standard Raman response.

\emph{Formulation.}--- A Raman process can be
either resonant or nonesonant. For the former the associated
electronic excitations are real, while they are virtual for the
latter. Therefore, resonant processes involve electronic transitions
at eV scale, the average visible photon
energy~\cite{Martin1983,Shvaika2004}. Intuitively, they are less
likely to affect low-energy spectra at $\Om \sim 10$ meV. Instead,
in this frequency range the nonresonant processes, which involve all
the low-energy excitations, should dominate. Thus, we start from the
standard description of nonresonant Raman response. 

Ignoring geometric pre-factors, the scattering intensity is given by~\cite{Devereaux07}
\begin{align}
I (\Om) =(1/Z) \sum_{I, F} e^{-\be E_I} \left| M_{F, I} \right|^2 \dl (E_I - E_F + \Om),
\nonumber
\end{align}
where $Z$ is the partition function, $\be$ is inverse temperature,
$\Om = \om_i - \om_f$ is the Raman shift, $\om_{i, f}$ are the in and out photon frequencies with polarizations
$\hat{e}_{i, f}$, and $E_{I, F}$ are the energies of the initial and final states.
We assume that the cutoff of the low-energy effective theory is
$D \ll \bar{\om}$, where $\bar{\om} = (\om_i + \om_f)/2$. In most cases of interest $D/\bar{\om} \sim$
0.1 or less. Then, the matrix element $M_{F, I}$ can be approximated by
[for details see the Supplementary Material (SM), Sec.~A in Ref.~\onlinecite{si}]
\begin{align}
\label{eq:Mfi}
M_{F, I} \approx \sum_{\al, \be} (e_{i, \al} e^{\ast}_{f, \be} )
\la F | \left( \hv_{\al \be} - \left[ \hv_{\al} \, , \, \hv_{\be} \right]/\bar{\om} \right) | I \ra.
\end{align}
Here $\hv_{\al} \equiv \sum_{\bk} \ptl \ham_{\bk}/\ptl k_{\al}$ and
$\hv_{\al \be} \equiv \sum_{\bk} \ptl^2 \ham_{\bk}/(\ptl k_{\al} \ptl k_{\be})$
are the paramagnetic and diamagnetic currents, respectively, and $\ham_{\bk}$ is the
non-interacting Hamiltonian at wavevector $\bk$.

We consider a quasi two-dimensional system for which $I_{x_{\tht} y_{\tht}}(\Om)$ is the Raman intensity
for in and out polarizations along $\hx_{\tht}$ and $\hy_{\tht}$, respectively, with
$\hx \cdot \hx_{\tht} = \cos \tht$, as shown in Fig.~\ref{fig1}(a). The angle dependent antisymmetric Raman intensity
can be expressed as (see SM, Sec.~B for details~\cite{si})
\begin{align}
\label{eq:IAtheta}
I_A (\Om, \tht) = 2 \cos (2 \tht) I_{A_{2g}}^{B_{2g}} (\Om) - 2 \sin (2\tht) I_{A_{2g}}^{B_{1g}} (\Om),
\end{align}
where
\begin{align}
\label{eq:I-partial}
&I_{A_{2g}}^{ i} (\Om) \equiv (1/Z) \sum_{I, F} e^{-\be E_I}  \dl (E_I - E_F + \Om)
\nonumber \\
&\times
\left[ \la I | \hGa_{A_{2g}} | F \ra \la F | \hGa_i | I \ra
- \la I | \hGa_i | F \ra \la F | \hGa_{A_{2g}} | I \ra \right],
\end{align}
with $i = (B_{1g}, B_{2g})$. The Raman operators $\hGa_{A_{2g}} \equiv
\left[ \hv_x , \hv_y \right]/\bar{\om}$~\cite{Kashuba09,Burdin20,Riccardi19},
$\hGa_{B_{1g}} \equiv (\hv_{xx} - \hv_{yy})/2$ and $\hGa_{B_{2g}} \equiv \hv_{xy}$.
Note, $\hGa_{A_{2g}}$ is an antisymmetric rank two tensor and is
an antihermitian operator, while $\hGa_i$ are symmetric rank two tensors and Hermitian operators.

Two comments are in order. First, the commutator $\hGa_{A_{2g}}$ is non-zero only for
multi-orbital systems, and it cannot be constructed within the widely-used effective mass approximation
that neglects the paramagnetic current~\cite{Devereaux07}. Moreover,
antihermiticity enforces that $\hGa_{A_{2g}}$ has only interband terms of the type
$(\cda_{\bk, 1} c_{\bk, 2} -\cda_{\bk, 2} c_{\bk, 1})$ between bands $(1, 2)$. This ensures that
antisymmetric responses have no intraband contribution.
Second, since $A_{2g} \otimes B_{2g} = B_{1g}$, finite $I_{A_{2g}}^{B_{2g}} (\Om)$
implies broken reflection symmetries
$M_{x^{\prime}}: (x, y) \rightarrow (y, x)$ and $M_{y^{\prime}}: (x, y) \rightarrow (-y,- x)$,
where $(x^{\prime}, y^{\prime})$ are
$\pi/4$ rotated counterclockwise with respect to $(x, y)$. Likewise, finite
$I_{A_{2g}}^{B_{1g}} (\Om)$ implies broken reflections
$M_{x}: (x, y) \rightarrow (x, -y)$ and $M_{y}: (x, y) \rightarrow (-x, y)$. Thus,
even though we use tetragonal notation,
antisymmetric Raman response is finite only for orthorhombic or lower symmetry systems.

\emph{Fluctuation dissipation relation.}---
The next step is to identify the susceptibilities that can be extracted from $I_A (\Om, \tht)$. For this we define the
imaginary time ordered function
$
\chi_{A_{2g}}^i (\tau) \equiv - \la T_{\tau}  \hGa_{A_{2g}} (\tau)  \hGa_i (0) \ra,
$
over the interval $\tau = [-\be , \be]$, and its Fourier transform by
$
\chi_{A_{2g}}^i (\tau) = (1/\be) \sum_{\Om_n} \chi_{A_{2g}}^i (i \Om_n) e^{-i \Om_n \tau}$
in terms of bosonic Matsubara frequencies $\Om_n$. Here $\la \cdots \ra$ implies thermodynamic average.
We also define the related cross-susceptibility
$\phi_{A_{2g}}^i (i \Om_n) \equiv \chi_{A_{2g}}^i (i \Om_n) - \chi_{A_{2g}}^i (-i \Om_n)$.
Note, by definition, $\chi_{A_{2g}}^i (-i \Om_n) = \chi_i^{A_{2g}} (i \Om_n)$.
Using Lehmann representation
we get (see SM, Sec.~C for details~\cite{si})
\begin{align}
\label{eq:phi}
\phi_{A_{2g}}^i (i \Om_n)  &= \frac{1}{Z} \sum_{n, m}
\left[ ( \hGa_{A_{2g}})_{nm} (\hGa_i)_{mn} -  (\hGa_i)_{nm} ( \hGa_{A_{2g}})_{mn} \right]
\nonumber \\
&\times
(e^{-\be E_n} - e^{\be E_m})/(i \Om_n + E_n - E_m),
\end{align}
where $(\hat{O})_{nm} \equiv \la n | \hat{O} | m \ra$. Comparing Eqs.~\eqref{eq:I-partial} and \eqref{eq:phi}
we obtain the fluctuation-dissipation relation
\begin{align}
\label{eq:fluc-dissp}
I_{A_{2g}}^{ i} (\Om)  = - (1/\pi) [1 + n_B(\Om)] {\rm Im} \phi_{A_{2g}}^i (\Om + i\eta),
\end{align}
where $n_B(\Om) \equiv 1/(e^{\be \Om} -1)$ is the Bose function.
By contrast, once symmetry resolved, standard Raman response gives the auto-susceptibility
${\rm Im}\chi_j^j(\Om + i\eta)$, where
$
\chi_j^j (\tau) \equiv - \la T_{\tau}  \hGa_j (\tau)  \hGa_j (0) \ra,
$
and $j = (A_{1g}, B_{1g}, B_{2g}, A_{2g})$. Consequently, it cannot detect reflection symmetry breaking.

\emph{General properties.}---
We now compare the properties of $\phi_{A_{2g}}^i (i \Om_n) $  extracted from antisymmetric Raman response
with those of $\chi_j^j(i\Om_n)$.

(i) Eq.~\eqref{eq:phi} implies $\phi_{A_{2g}}^i (\Om + i\eta)^{\ast} = \phi_{A_{2g}}^i (\Om - i\eta)$,
using the antihermiticity of $\hGa_{A_{2g}}$. Moreover, $\phi_{A_{2g}}^i (i \Om_n)$ is odd in $\Om_n$.
Thus, the real and the imaginary parts of $\phi_{A_{2g}}^i (\Om + i\eta)$ are
odd and even functions of $\Om$, respectively. It is opposite in case of $\chi_j^j(\Om + i\eta)$, as shown in
Fig.~\ref{fig1}(c, d).

(ii) From the detailed balance relation~\cite{Loudon78}
$
 I_{y_{\tht} x_{\tht}}(-\Om) = e^{-\be \Om} I_{x_{\tht} y_{\tht}}(\Om)
$ (see SM, Sec.~D for details~\cite{si}),
we get $I_A (\Om \rightarrow 0, \tht)  \rightarrow 0$. Since $n_B(\Om) \propto 1/\Om$ in this limit,
we have
$\lim_{\Om \rightarrow 0} {\rm Im} \phi_{A_{2g}}^i (\Om + i\eta) \sim |\Om|^{\ga}$ with $\ga > 1$.
As discussed later, in practice ${\rm Im} \phi_{A_{2g}}^i (\Om + i\eta)$ from electronic excitations
[one loop graph in Fig.~\ref{fig1}(b)] is zero below a threshold, as shown in Fig.~\ref{fig1}(c). Under suitable
conditions a collective mode [two loop graph in Fig.~\ref{fig1}(b)] can appear inside this spectroscopic ``gap''.

(iii) From Eq.~\eqref{eq:phi} 
and the behavior of $\phi_{A_{2g}}^i (\Om)$ at large $\Om$
we obtain (see SM, Sec.~E for details~\cite{si})
\begin{align}
\label{eq:sum-rule}
\int_{-\infty}^{\infty} \frac{d \Om}{2\pi}  \phi_{A_{2g}}^i (\Om + i\eta)
= i \la [\hGa_i \, , \,  \hGa_{A_{2g}}] \ra.
\end{align}
Since the right hand side is zero for a tetragonal system, it acts
as an order parameter for reflection symmetry breaking which can be extracted from spectroscopy.
By contrast, in standard Raman the frequency integral of
${\rm Im}\chi_j^j(\Om + i\eta)/\Om$ is related to a static susceptibility by Kramers-Kronig
relation~\cite{Gallais16,Gallais13}.
Note, Eq.~\eqref{eq:sum-rule} is not a sum rule in the spirit of the $f$-sum rule, since the Raman operators
are not related to any conservation law. Thus, the above integral is typically a function of temperature.

(iv) Reciprocity relation: We define the correlation function
$C_{A_{2g}}^i(t) \equiv \langle \hGa_{A_{2g}}(t) \hGa_i (0) \rangle$
and $C_{A_{2g}}^i(\Om) \equiv \int_{-\infty}^{\infty} dt \exp (i \Om
t) C_{A_{2g}}^i(t)$, and likewise $C_i^{A_{2g}}(t)$ and
$C_i^{A_{2g}}(\Om) $. The antisymmetric Raman intensities are given
by $I_{A_{2g}}^i (\Om) = C_{A_{2g}}^i(\Om)  - C_i^{A_{2g}}(\Om) $.
We consider a time reversal symmetric system in an external magnetic
field ${\bf h}$. As the Raman operators are even under time
reversal, and $\hGa_{A_{2g}}$ is antihermitian, one can show that
$C_{A_{2g}}^i(t, h) = - C_i^{A_{2g}}(t, -h)$. This gives the
\emph{reciprocity relation} (see SM, Sec.~F for details~\cite{si}) 
\beq
\label{eq:reciprocity} 
I_{A_{2g}}^i (\Om, h) = I_{A_{2g}}^i (\Om, -h) . 
\eeq 
A violation of the above was reported recently for the
CDW phase of GdTe$_3$~\cite{Wulferding24} .

Analogous to antisymmetric response,
cross-susceptibilities $B_{1g} \otimes B_{2g}$ or $E_{g}^1 \otimes E_{g}^2$
are extracted by subtracting right-left
circular polarizations signal from the left-right one in reference tetragonal or hexagonal systems,
respectively~\cite{Liu22,Perfetti24,Arita25}.
However, these do not involve the $A_{2g}$-operator, and hence
do not have the features discussed here.

\emph{Examples.}---
Having discussed the general properties of $\phi_{A_{2g}}^i$, we illustrate them with
examples from systems of current interest.

\emph{(1) Electronic continuum.}---
This example is taken to illustrate the antisymmetric Raman response of a normal metal.
We consider a tight binding Hamiltonian of $(p_x, p_y)$ orbitals that is often studied in the context of the
tritellurides~\cite{Yao06,Eiter13,Cano24}. It is given by $\ham_0 = \sum_{\bk} h_{\bk}$,
\[
h_{\bk} = \sum_{\nu = x, y} \ep_{\bk, \nu} \cda_{\bk, \nu} c_{\bk, \nu}
+ V_{\bk} \cda_{\bk, x} c_{\bk, y} + {\rm h.c.},
\]
with $\ep_{\bk, x} = -2 t \cos k_x + 2 t^{\prm} \cos k_y -\mu$,
$\ep_{\bk, y} = \ep_{\bk, x} (k_x \leftrightarrow k_y)$, and
$V_{\bk} = -2 V_0 \sin k_x \sin k_y - \ep_m \cos k_x \cos k_y$. 
The $\ep_m$-term breaks $(M_x, M_y)$ symmetries, while $(M_{x^{\prm}},
M_{y^{\prm}})$ are intact, making the system orthorhombic with
finite $\phi_{A_{2g}}^{B_{1g}}(\Om)$ and zero
$\phi_{A_{2g}}^{B_{2g}}(\Om)$. 
A non-zero $\hGa_{A_{2g}}$ requires finite orbital hybridization. Thus,
its bare scale is set by $(V_0, \ep_m)$, while that of $\hGa_{B_{1g}}$ by $(t, t^{\prm})$.

For simplicity in the rest of this work we
set temperature $T=0$, since finite-$T$ corrections are small at low
temperatures (for details see SM, Sec.~G~\cite{si}). Also, we set $\bar{\om} = t$.
In Fig.~\ref{fig1}(c, d) we compare
$\phi_{A_{2g}}^{B_{1g}}(\Om)$ and $\chi_{B_{1g}}^{B_{1g}}(\Om)$ 
for a standard set of tight binding parameters.
Their overall scales reflect the ratio $V_0/t$. 
Otherwise, they have  two crucial differences.
First, the absence of
intraband Drude contribution in ${\rm
Im}\phi_{A_{2g}}^{B_{1g}}(\Om)$, which is imposed by
$\hGa_{A_{2g}}$. Second, the same is non-zero only above a threshold
frequency $\Om_{th}$. This is typical of orthorhombic (and lower
symmetry) systems where all band degeneracies are lifted (the
irreducible representations of the point group are one dimensional).
This implies that interband particle-hole excitations need a minimum
energy $\Om_{th} = 2V_0$ in this model. Thus the antisymmetric
response has a spectroscopic ``gap'' even though the system is
metallic and, in principle, one can observe a collective mode of a
broken symmetry phase in this gap. This complete
absence of Drude response, even in a good metal,
is a matrix element effect, distinct from the partial suppression
seen in standard Raman channel
$\chi_{A_{1g}}^{A_{1g}}$~\cite{Freericks2001}. The latter is due to
a finite overlap between the $A_{1g}$ operator and $\hat{N}$, the
total particle number. Since $\hat{N}$ is conserved, there is a
partial cancellation in $\chi_{A_{1g}}^{A_{1g}}$
but, being partial, it does not reveal the scale  $\Om_{th}$.

\begin{figure}[t]
\begin{center}
\includegraphics[width=0.45\textwidth]{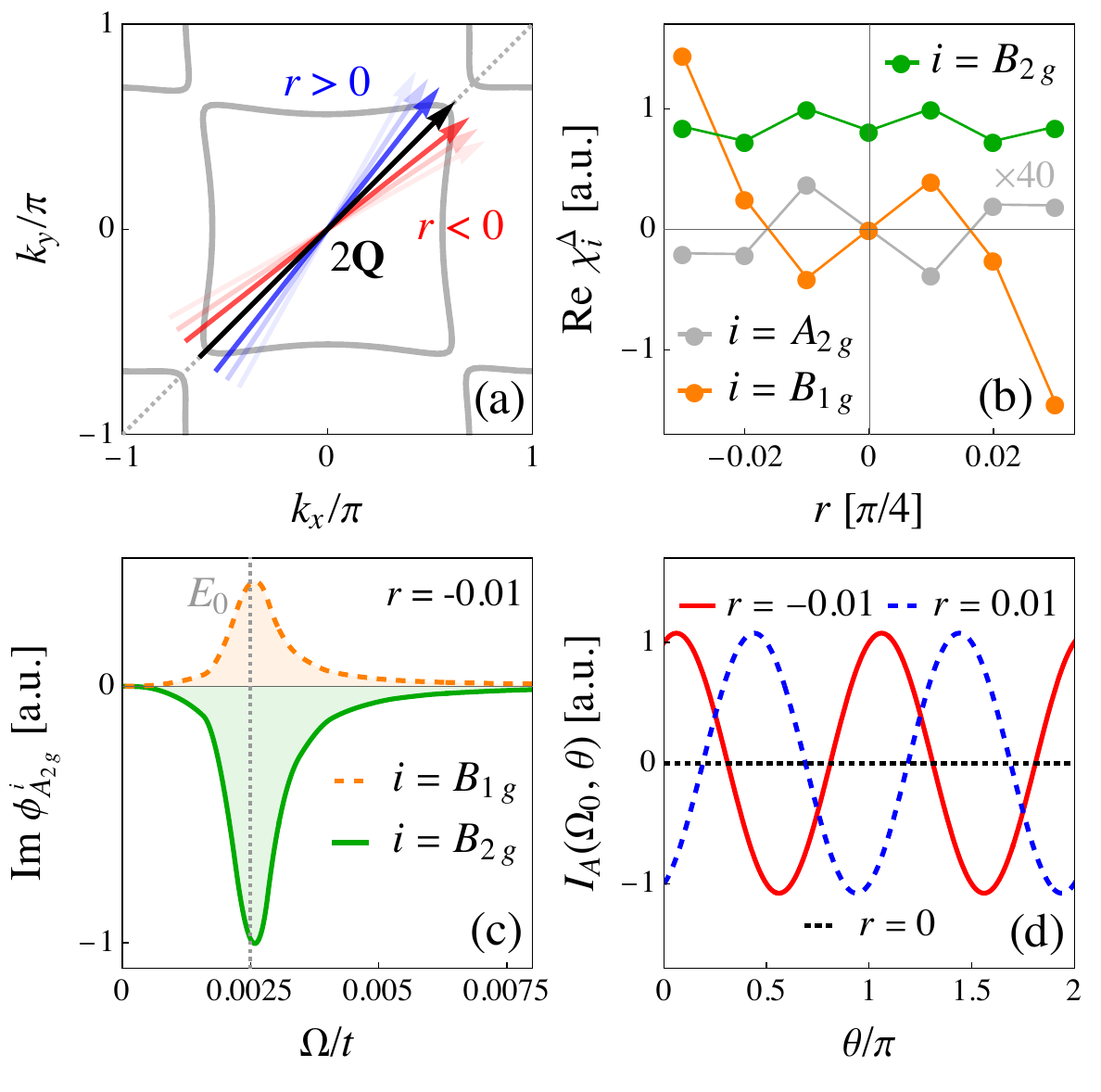}
\caption{(Color Online) (a) Fermi surface of the tritellurides for $T > T_c$. $2\bQ$ is the CDW wavevector. It is tilted by
an angle $r$ compared to the direction $x = y$.
(b) Dependence of various susceptibilities as a function of $r$.  By symmetry,
${\rm Re} \chi_{A_{2g}}^{\Delta}(E_0) = {\rm Re} \chi_{B_{1g}}^{\Delta}(E_0) = 0$ for $r=0$.
A finite $r$ is necessary to explain the monoclinicity of the $T < T_c$ phase. (c) Antisymmetric Raman responses due to the
amplitude mode of the CDW. The absolute signs of the two responses and their relative strengths are consistent with the
experiments. (d) Angular variation of the antisymmetric Raman intensity, and its dependence on the sign of $r$.
}
\label{fig2}
\end{center}
\end{figure}

\emph{(2) Amplitude mode of CDW.}---
We explain how the collective mode of a CDW, with energy $E_0$, appears in an antisymmetric channel, as recently
reported for the tritellurides~\cite{Burch22}. We also discuss the microscopic
information that can be extracted from such observation.

All the available Raman studies~\cite{Burch22,Wulferding24,Gallais25}
on the tritellurides show similar collective mode features.
Using notations developed here they can be summarized as follows.
(i) $ I_A(E_0, 0) > 0$, which implies ${\rm Im} \phi_{A_{2g}}^{B_{2g}}(E_0) < 0$;
(ii) $I_A(E_0, \pi/4) < 0$, implying ${\rm Im} \phi_{A_{2g}}^{B_{1g}}(E_0) > 0$; and
(iii) $I_A(E_0, 0) \gg |I_A(E_0, \pi/4)|$, which gives
 $|{\rm Im} \phi_{A_{2g}}^{B_{2g}}(E_0) | \gg {\rm Im} \phi_{A_{2g}}^{B_{1g}}(E_0) > 0$.
 Our goal is to model these features and draw conclusions about the CDW state.

 The coupling of light to a charge neutral CDW amplitude mode involve two fermion
 loops~\cite{Klein82,Lee74,Cea14,Udina19}, as shown in Fig.~\ref{fig1}(b).
 Since each loop is nonzero, all four mirror symmetries of a reference tetragonal system have to be broken. Thus,
 observing the collective mode implies \emph{monoclinicity}. 
 The tritellurides are orthorhombic to begin with broken mirrors
 $(M_x, M_y)$ in the normal state at temperature $T > T_c$, the CDW transition temperature.
 Thus, the CDW breaks  the mirrors $(M_{x^{\prime}}, M_{y^{\prime}})$ along with lattice translation
 symmetry~\cite{Cano24}.
We postulate that the former is due to a tilt of the ordering wavevector $2\bQ$ from the high-symmetry direction $x=y$,
at an angle $(1+r)\pi/4$ from the $x$-axis,
with $r$ the tilt parameter as shown in Fig.~\ref{fig2}(a)  along with the normal state Fermi surface. Note, the precise origin of the monoclinicity is not important for what follows, and other possibilities have
been discussed~\cite{Cano24, Burch24}.

Experimentally, $E_0$ is about two orders of magnitude smaller than the tight-binding
scales~\cite{Brouet04,Brouet08,Moore10,Bolloch24}, and it lies well inside the
interband ``gap'' discussed in example ($1$). Thus, ${\rm Im} \chi_{A_{2g}}^{\Delta}(E_0) = 0$, and
\beq
\label{eq:2-bubble}
{\rm Im} \phi_{A_{2g}}^i (\Om \sim E_0) \approx 2 [{\rm Re} \chi_{A_{2g}}^{\Delta}(\Om)][{\rm Re}
\chi_{\Delta}^{i}(\Om)] {\rm Im} D(\Om),
\eeq
where $D(\Om) = 2E_0/[(\Om + i\ga_0)^2 - E_0^2]$ is the amplitude mode propagator,
with $\ga_0$ its broadening.
This relation applied to the experimental implication (iii) discussed above gives
$|[{\rm Re} \chi_{\Delta}^{B_{2g}}(E_0)]| \gg |[{\rm Re} \chi_{\Delta}^{B_{1g}}(E_0)]|$.
In turn, it indicates that the CDW order parameter
is predominantly of the inter-orbital type, with an $xy$ or $B_{2g}$ character
in the orbital space $(p_x, p_y)$.

The above suggests that the CDW phase can be described by the mean field Hamiltonian
$
\ham = \ham_0 + \Delta \sum_{\bk} [ \cda_{\bk + \bQ, x} c_{\bk - \bQ, y} +  \cda_{\bk + \bQ, y} c_{\bk - \bQ, x} + {\rm h.c.}],
$
where $\ham_0$ is that of example ($1$), and $\Delta$ is the CDW potential with $2\bQ$ ordering wavevector.
The amplitude mode describes fluctuations of $\Delta$, hybridized with the phonon that triggers the CDW
instability~\cite{Cea14,Udina19,Maschek15,Grasset18}. This hybridization implies that $E_0 \neq \Delta$.
The computational details are given in the SM, Sec.~H~\cite{si}.
Note, ${\rm Re} \chi_i^{\Delta}(E_0) = {\rm Re} \chi_{\Delta}^i(E_0)$ for $i = (B_{1g}, B_{2g})$, while
${\rm Re} \chi_{A_{2g}}^{\Delta}(E_0) = - {\rm Re} \chi_{\Delta}^{A_{2g}}(E_0)$.
 In Fig.~\ref{fig2}(b) we show the tilt angle $r$
dependence of ${\rm Re} \chi_i^{\Delta}(E_0)$ for $i = (A_{2g}, B_{1g}, B_{2g})$.
Since $\Delta$ is a $B_{2g}$ variable for
$r=0$, we get ${\rm Re} \chi_{A_{2g}}^{\Delta}(E_0) = {\rm Re} \chi_{B_{1g}}^{\Delta}(E_0) = 0$ in this limit. By contrast,
${\rm Re} \chi_{B_{2g}}^{\Delta}(E_0)$ is nearly independent of $r$.

In Fig.~\ref{fig2}(c) we show the amplitude mode in the two antisymmetric
Raman channels with their absolute signs and relative magnitudes consistent with the experiments.
Since $\phi_{A_{2g}}^{B_{2g}}$ is a $B_{1g}$ quantity, it is proportional to $r$, and the
fact ${\rm Im} \phi_{A_{2g}}^{B_{2g}}(E_0) < 0$
(the monoclinic domains are pinned to the orthorhombic ones)
is consistent with $r < 0$ in the model.
We find that the remaining two properties, namely ${\rm Im} \phi_{A_{2g}}^{B_{1g}}(E_0) > 0$ and
$|{\rm Im} \phi_{A_{2g}}^{B_{2g}}(E_0) | \gg {\rm Im} \phi_{A_{2g}}^{B_{1g}}(E_0)$
depend on microscopic details, and they could be obtained only by 
fine tuning parameters such as the chemical potential $\mu$ and the term $\ep_m$ in $\ham$.
This is because $\mu$ changes considerably the Fermi surface in the CDW state, which in turn affects
the low-energy particle-hole excitations. Likewise, $\ep_m$ breaks $B_{2g}$ symmetry and, thus, it
modulates the relative weights between the peaks of Fig.~\ref{fig2}(c).
Finally, Fig.~\ref{fig2}(d) shows the two-fold symmetric
angular dependence of the antisymmetric Raman intensity as defined by Eq.~\eqref{eq:IAtheta}.
The two cases $r < 0$ and $r > 0$ are $\pi/2$
phase shifted because, with sign change of $r$, the sign of ${\rm Im} \phi_{A_{2g}}^{B_{2g}}(E_0)$ changes
but not that of ${\rm Im} \phi_{A_{2g}}^{B_{1g}}(E_0)$.

Thus, signals of collective modes in antisymmetric Raman channels yield several useful microscopic properties
that cannot be accessed otherwise.
They will be useful to improve the tight binding model, 
and to estimate the relevance of their multi-layer and non-symmorphic nature.
\begin{figure}[t]
\begin{center}
\includegraphics[width=0.45\textwidth]{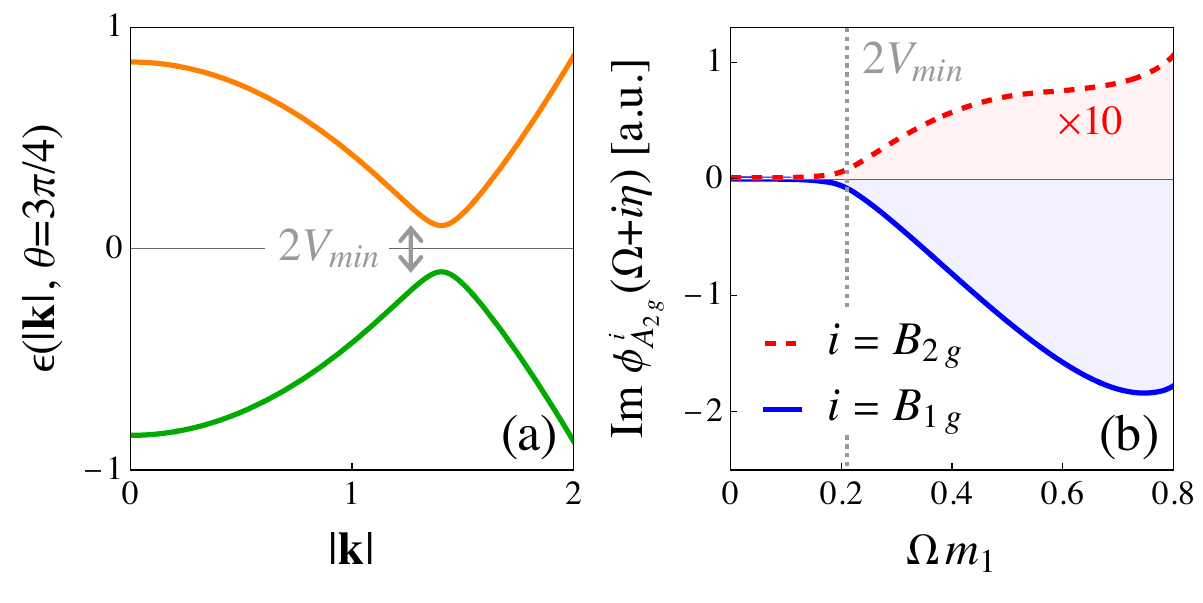}
\caption{(Color Online)
(a) Band dispersions of an excitonic insulator in polar coordinates, with minimum charge gap of 2$V_{\rm min}$.
(b) Antisymmetric Raman spectra in this phase with $\Om_{th} = 2V_{\rm min}$.
Integrated weight of ${\rm Im} \phi_{A_{2g}}^{B_{1g}} (\Om)$ measures electronic monoclinicity.
}
\label{fig3}
\end{center}
\end{figure}

\emph{(3) Exciton insulator.}---
Ta$_2$NiSe$_5$, which has attracted considerable interest in recent times, undergoes a
transition from a high-$T$ orthorhombic metal to a low-$T$ monoclinic insulator due to the formation of excitons, or
particle-hole pairs~\cite{Salvo86,Lu17,Mazza20,Ohta25}.
Our goal is to illustrate that antisymmetric Raman response will be an interesting probe  of this
and other similar insulating systems.

We model the metallic phase by two bands $\pm \ep_{\bk}$ with $\ep_{\bk} = k_x^2/(2m_1) + k_y^2/(2m_2) - E_0$,
with $m_1 \neq m_2$ that
breaks mirrors $(M_{x^{\prime}}, M_{y^{\prime}})$. In this phase $\phi_{A_{2g}}^{B_{1g}}$ vanishes,
while $\phi_{A_{2g}}^{B_{2g}}$ is nonzero (zero in this simplified model). In the insulating
phase the bands have finite hybridization $V_{\bk} = (k_x k_y + c)/m$, which opens a gap and
breaks the mirrors $(M_x, M_y)$ as well, making $\phi_{A_{2g}}^{B_{1g}}$ non-zero.
We show its imaginary part in
Fig.~\ref{fig3},
where $\Om_{th} \sim 2V_{\rm min}$ is now a measure of the charge gap. Furthermore, since lattice strain
effects are absent in Raman response~\cite{Gallais16,Gallais16b},
the electronic monoclinicity can be extracted
from the integrated spectrum using  Eq.~\eqref{eq:sum-rule}.

\emph{Conclusion.}--- Antisymmetric Raman response measures
cross-susceptibilities of two Raman operators, of which one is $\hGa_{A_{2g}}$
that filters out intraband excitations. The spectra have ``gaps'' below a threshold which
measures minimum interband excitation energy in metals and the charge gap in insulators. It can detect
subtle mirror symmetry breaking such as CDW transitions in tritellurides and excitonic insulator
transition in Ta$_2$NiSe$_5$. In symmetry broken phases a
collective mode can appear inside the gap which has
microscopic information such as the orbital character of the order parameter.
Thus, antisymmetric Raman response
is a powerful tool to probe low symmetry materials, both metallic and insulating.

\emph{Acknowledgement.}---
We are thankful to T. Freitas, Y. Gallais, A. Meszaros, and A. Sacuto for insightful discussions.
We acknowledge financial support from French
Agence Nationale de la Recherche (ANR) grant ANR-23-CE30-0030 (SUPER2DTMD).


\appendix

\renewcommand{\figurename}{{\bf Supplementary Figure}}
\renewcommand{\thefigure}{S\arabic{figure}}
\setcounter{figure}{0}

\renewcommand{\tablename}{{\bf Supplementary Table}}
\renewcommand{\thetable}{S\arabic{table}}
\setcounter{table}{0}

\renewcommand\theequation{S\arabic{equation}}
\setcounter{equation}{0}

\setcounter{page}{1}

\section*{Supplementary Information}

\subsection{Derivation of Eq.~(1)}

In general, the Raman matrix element is given by
\begin{align}
\label{eq:matrixelement}
M_{F, I} &= \sum_{\al, \be} (e_{i, \al} e^{\ast}_{f, \be} )
\left[ \la F | \hv_{\al \be} | I \ra
+ \sum_M \frac{\la F | \hv_{\al} | M \ra \la M | \hv_{\be} | I \ra}{E_I - E_M - \om_f} \right.
\nonumber \\
&+ \left. \frac{\la F | \hv_{\be} | M \ra \la M | \hv_{\al} | I \ra}{E_I - E_M + \om_i} \right],
\end{align}
where $\Om = \om_i - \om_f$ is the Raman shift, $\om_{i, f}$ are the in and out photon frequencies with polarizations
$\hat{e}_{i, f}$.
The paramagnetic and diamagnetic current operators, defined in the main text, are taken at zero external wavevector ($\bq =0$).
In other words, the photon momentum is neglected in the scattering process. This is usually a good approximation to describe Raman
response because the wavelength of visible light is much longer than the lattice spacing. As a result the response is dominated by
$\bq =0$ processes (vertical transitions), and finite-$\bq$ effects are negligible.

We assume that the cutoff of the effective theory  $D \ll \om_i, \om_f$. Since $(E_I - E_M) < D$, we can neglect it in
the denominators, and the intermediate states $M$ can be summed to unity. Writing
$\om_{i, f} = \bar{\om} \pm \Om/2$, and neglecting $\Om/\bar{\om}$ we get Eq.~(1) of the main text.

Note, in case of a resonant Raman scattering the
electronic excitations are real energy conserving processes. They
satisfy either $E_M = E_I + \om_i$ (resonant photon absorption) or
$\om_f = E_I - E_M$ (resonant emission), in which case one of the
energy denominators in Eq~\eqref{eq:matrixelement} vanish formally,
and the above approximation is invalid. However, resonant processes
invariably involve electronic excitations at eV energy scale, the
average visible photon energy. Intuitively, from separation of
energy scales argument, they are less likely to affect low-energy
spectra at $\Om \sim$ 10 meV. Instead, in this frequency range the
nonresonant processes, which involve all the low-energy excitations,
should dominate. 

\subsection{Derivation of Eqs.~(2) and (3)}
We consider the scattering geometry where the in-coming photon polarization is
$\hat{e}_i = \hx_{\tht} = \cos (\tht) \hx + \sin (\tht) \hy$, and the out-going photon polarization is
$\hat{e}_f = \hy_{\tht} = -\sin (\tht) \hx + \cos (\tht) \hy$. The associated Raman matrix element
can be written
\beq
M_{F, I} (\tht) = \la F | [ \hat{D}(\tht) + \hat{P} (\tht) ] | I \ra,
\eeq
where the diamagnetic term is
\[
\hat{D}(\tht) = \cos (2\tht) \hGa_{B_{2g}} - \sin (2\tht) \hGa_{B_{1g}},
\]
and the paramagnetic term is
\[
\hat{P} (\tht) = - \hGa_{A_{2g}},
\]
with $\hGa_{A_{2g}} \equiv  \left[ \hv_x , \hv_y \right]/\bar{\om}$,
$\hGa_{B_{1g}} \equiv (\hv_{xx} - \hv_{yy})/2$ and $\hGa_{B_{2g}} \equiv \hv_{xy}$.
We need to compare this scattering geometry with the one obtained by interchanging the
in and out polarization directions. Denoting the Raman matrix element of the latter by
$M_{F, I} (\tht_{\perp}) $, we get
\beq
M_{F, I} (\tht_{\perp}) = \la F | [ \hat{D}(\tht) - \hat{P} (\tht) ] | I \ra.
\eeq
Thus, the antisymmetric Raman response is
\begin{align}
I_A (\Om, \tht) =& I_{x_{\tht} y_{\tht}}(\Om) - I_{y_{\tht} y_{\tht}}(\Om)
\nonumber \\
=& \frac{1}{Z} \sum_{I, F} e^{-\be E_I}  \dl (E_I - E_F + \Om)
\nonumber\\
\times&
\left[ |M_{F, I} (\tht) |^2 - |M_{F, I} (\tht_{\perp}) |^2 \right].
\end{align}
Noting that $\hat{P} (\tht)$ is antihermitian, from the above we get Eqs.~(2) and (3)
of the main text.

\subsection{Derivation of Eq.~(4)}
The imaginary time ordered function
$
\chi_{A_{2g}}^i (\tau) \equiv - \la T_{\tau}  \hGa_{A_{2g}} (\tau)  \hGa_i (0) \ra,
$
can be expressed in the Lehmann basis as
\begin{align}
\chi_{A_{2g}}^i (\tau) &= - \tht(\tau) \frac{1}{Z} \sum_{n, m} e^{-\be E_n} (\hGa_{A_{2g}})_{nm} ( \hGa_i)_{mn}
e^{\tau(E_n - E_m)}
\nonumber\\
&- \tht(-\tau) \frac{1}{Z} \sum_{n, m} e^{-\be E_n}( \hGa_i)_{nm} (\hGa_{A_{2g}})_{mn}
e^{\tau(E_m - E_n)},
\end{align}
where $\tau$ is defined over the interval $[-\be, \be]$. From the above we deduce that
$\chi_{A_{2g}}^i (\tau > 0) = \chi_{A_{2g}}^i (\tau - \be < 0)$ and, in particular, $\chi_{A_{2g}}^i (\be) = \chi_{A_{2g}}^i(0)$.
The periodic boundary condition implies that $\chi_{A_{2g}}^i (\tau)$ can be expanded in bosonic Matsubara frequency.
We get
\begin{align}
\label{eq:chi}
\chi_{A_{2g}}^i (i \Om_n) &= \int_0^{\be} d \tau e^{i \Om_n \tau}  \chi_{A_{2g}}^i (\tau)
\nonumber\\
&= \frac{1}{Z} \sum_{n, m} (\hGa_{A_{2g}})_{nm} ( \hGa_i)_{mn} \,
\frac{e^{-\be E_n} - e^{-\be E_m}}{i\Om_n +E_n - E_m}.
\end{align}
This implies that
\begin{align}
\label{eq:phi}
\phi_{A_{2g}}^i (i \Om_n)
&= \chi_{A_{2g}}^i (i \Om_n)  - \chi_{A_{2g}}^i (-i \Om_n)
\nonumber\\
&= \frac{1}{Z} \sum_{n, m}
\left[ ( \hGa_{A_{2g}})_{nm} (\hGa_i)_{mn} -  (\hGa_i)_{nm} ( \hGa_{A_{2g}})_{mn} \right]
\nonumber \\
&\times
\frac{e^{-\be E_n} - e^{\be E_m}}{i \Om_n + E_n - E_m},
\end{align}
which is Eq.~(4) of the main text.

\subsection{Comparing Stokes and anti-Stokes processes}
We consider a Raman scattering process where the in-photon has frequency and polarization
$\om_1$ and $\hat{\al}$, respectively, while for the out-photon they are $\om_2$ and $\hat{\be}$, respectively.
We take $\Om = \om_1 - \om_2 > 0$, i.e., this is a Stokes process. The scattering intensity for this process is
given by
\begin{align}
I_{\al \be} (\om_1, \om_2) =  \frac{1}{Z} \sum_{i, f, m} e^{-\be E_i} \dl(E_i - E_f + \Om) \left|
M_{fi}^{\al \be} (\om_1, \om_2)\right|^2,
\nonumber
\end{align}
where
\begin{align}
\label{eq:M-stoke}
M_{fi}^{\al \be} (\om_1, \om_2) = (\hv_{\al \be})_{fi}
+ \frac{(\hv_{\al})_{fm} (\hv_{\be})_{mi}}{E_{im} - \om_2}
+ \frac{(\hv_{\be})_{fm} (\hv_{\al})_{mi}}{E_{im} + \om_1},
\end{align}
with $E_{im} \equiv E_i - E_m$. This intensity is to be compared with an anti-Stokes process
where the in-photon has frequency and polarization $\om_2$ and $\hat{\be}$, respectively, while
those for the out-photon are $\om_1$ and $\hat{\al}$, respectively. The intensity of this process
is
\begin{align}
I_{\be \al} (\om_2, \om_1) =  \frac{1}{Z} \sum_{i, f, m} e^{-\be E_i} \dl(E_i - E_f - \Om) \left|
M_{fi}^{\be \al} (\om_2, \om_1)\right|^2,
\nonumber
\end{align}
where
\begin{align}
\label{eq:M-antistoke}
M_{fi}^{\be \al} (\om_2, \om_1) &= (\hv_{\be \al})_{fi}
+ \frac{(\hv_{\be})_{fm} (\hv_{\al})_{mi}}{E_{im} - \om_1}
+ \frac{(\hv_{\al})_{fm} (\hv_{\be})_{mi}}{E_{im} + \om_2}
\end{align}
Using energy conservation relation $E_i + \om_2 = E_f + \om_1$ the anti-Stokes matrix element can be re-written as
\begin{align}
\label{eq:M-relation}
&M_{fi}^{\be \al} (\om_2, \om_1) = (\hv_{\be \al})_{fi}
+ \frac{(\hv_{\be})_{fm} (\hv_{\al})_{mi}}{E_{fm} - \om_2}
+ \frac{(\hv_{\al})_{fm} (\hv_{\be})_{mi}}{E_{fm} + \om_1}
\nonumber\\
&=
\left[
(\hv_{\al \be})_{if}
+ \frac{(\hv_{\al})_{im} (\hv_{\be})_{mf}}{E_{fm} - \om_2}
+ \frac{(\hv_{\be})_{im} (\hv_{\al})_{mf}}{E_{fm} + \om_1}
\right]^{\ast}
\nonumber\\
&= \left[ M_{if}^{\al \be} (\om_1, \om_2) \right]^{\ast}
\end{align}
Using the above relation we get
\begin{align}
I_{\be \al} (\om_2, \om_1) &=
 \frac{1}{Z} \sum_{i, f, m} e^{-\be E_i} \dl(E_i - E_f - \Om) \left|
M_{if}^{\al \be} (\om_1, \om_2)\right|^2
\nonumber\\
&=
 \frac{1}{Z} \sum_{i, f, m} e^{-\be E_f} \dl(E_f - E_i - \Om) \left|
M_{fi}^{\al \be} (\om_1, \om_2)\right|^2
\nonumber\\
&= e^{-\be \Om} I_{\al \be}(\om_1, \om_2).
\end{align}
The above relation implies that for $\Om \rightarrow 0$, the antisymmetric Raman intensity $I_A(\Om, \tht) \rightarrow 0$.

\subsection{Derivation of Eq.~(6)}

The Lehmann representation of a standard Raman auto-susceptibility can be obtained from Eq.~\eqref{eq:chi}
by replacing the index $A_{2g} \rightarrow i$. In this case, for a complex frequency $s$ we get
\begin{align}
\lim_{|s| \rightarrow \infty} s \chi_i^i(s) &= \left[- \frac{1}{Z } \sum_{n, m} (\hGa_i)_{nm} (\hGa_i)_{mn}
\left( e^{-\be E_m} - e^{-\be E_n} \right) \right]
\nonumber \\
&= 0.
\end{align}
Since $\chi_i^i(s)$ falls faster than $1/s$ for large $s$, it implies that
\beq
\label{eq:sum-chi}
\int_{-\infty}^{\infty} \frac{d \om}{2\pi} \chi_i^i(\om + i\eta) = 0,
\eeq
since the frequency integral can be performed by closing the contour in the upper half complex frequency plane
where the function is analytic due to its causal structure.

For the antisymmetric cross-susceptibility the behavior is different.  From Eq.~\eqref{eq:phi} we have
\begin{align}
\lim_{|s| \rightarrow \infty} s \phi_{A_{2g}}^i(s) &= 2 \la [\hGa_{A_{2g}} \, , \,  \hGa_i] \ra,
\end{align}
which is non-zero, in general. Thus, the argument of closing the contour in the upper half plane is no longer valid.
Instead, we note that the real and imaginary parts of $\phi_{A_{2g}}^i(\om + i\eta)$ are odd and even, respectively. We get
\begin{align}
\int_{-\infty}^{\infty} \frac{d \om}{2\pi} \phi_{A_{2g}}^i(\om + i\eta) &= i \int_{-\infty}^{\infty} \frac{d \om}{2\pi} {\rm Im} \phi_{A_{2g}}^i(\om + i\eta)
\nonumber \\
&= i \la [\hGa_i \, , \,  \hGa_{A_{2g}}] \ra,
\end{align}
which is Eq.~(6) in the main text. Thus, the above relation is simply a consequence of $\phi_{A_{2g}}^i$ being a cross-susceptibility
of two non-commuting operators. Note, while the left side of the above relation can be deduced from spectroscopy, in principle,
the right side is a thermodynamic quantity.

\subsection{Reciprocity relation}
From the definition of the correlation functions $C_{A_{2g}}^i(t)$, $C_i^{A_{2g}}(t)$ and their Fourier transforms, we have
\begin{align}
C_{A_{2g}}^i(\Om) -& C_i^{A_{2g}}(\Om) =
\frac{1}{Z} \sum_{n, m} e^{-\be E_n}
\left[ (\hGa_{A_{2g}})_{nm} (\hGa_i)_{mn} \right.
\nonumber\\
&-  \left. (\hGa_i)_{nm} (\hGa_{A_{2g}})_{mn} \right]
\int_{-\infty}^{\infty} d t e ^{i(E_n - E_m + \Om)t}
\nonumber\\
&= I_{A_{2g}}^i (\Om)
\end{align}

Next, we establish how the correlation functions vary under time reversal operator $\hK$. The antilinearity of $\hK$ implies
the relation
\beq
\la n | \hat{A} | n \ra = \la \bar{n} | \hK \hat{A}^{\dagger} \hK^{-1} | \bar{n} \ra,
\eeq
where $\hat{A}$ is any operator and $| \bar{n} \ra = \hK | n \ra$. Since a complete set of states $\{ | n \ra \}$ can be
replaced by another complete set $\{ | \bar{n} \ra \}$, we have
\beq
{\rm Tr} (\hat{A}) = {\rm Tr} (\hK  \hat{A}^{\dagger} \hK^{-1}).
\eeq
We consider a system which is intrinsically time reversal symmetric such that the Hamiltonian $\hH$ satisfies
$\hK \hH (h) \hK^{-1} = \hH(-h)$, where $h$ is the external magnetic field. In the above equation we set
$ \hat{A} = e^{-\be \hH(h)} e^{i \hH(h) t} \hGa_{A_{2g}} e^{-i \hH(h) t} \hGa_i$. This gives
\begin{align}
&C_{A_{2g}}^i (t, h) =
\frac{1}{Z} {\rm Tr} \left( e^{-\be \hH(h)} e^{i \hH(h) t} \hGa_{A_{2g}} e^{-i \hH(h) t} \hGa_i \right)
\nonumber\\
&=
- \frac{1}{Z} {\rm Tr} \left( \hK \left[\hGa_i e^{i \hH(h) t} \hGa_{A_{2g}} e^{-i \hH(h) t}  e^{-\be \hH(h)}  \right]\hK^{-1} \right)
\nonumber\\
&=
- \frac{1}{Z} {\rm Tr} \left( \hGa_i e^{-i \hH(-h) t} \hGa_{A_{2g}} e^{i \hH(-h) t}  e^{-\be \hH(-h)}   \right)
\nonumber\\
&=
- \frac{1}{Z} {\rm Tr} \left( e^{-\be \hH(-h)}   e^{i \hH(-h) t} \hGa_i e^{-i \hH(-h) t}  \hGa_{A_{2g}} \right)
\nonumber\\
&= - C_i^{A_{2g}} (t, -h).
\end{align}
In the above we used the fact that the Raman operators are time reversal invariant and that $ \hGa_{A_{2g}} $ is
antihermitian.
Thus, the Fourier transforms satisfy the relation $C_{A_{2g}}^i (\Om, h) = - C_i^{A_{2g}} (\Om, -h)$, from which
Eq.~(7) of the main text follows.

\subsection{Example ($1$), antisymmetric response from electronic continuum}

We start from the single layer, two orbital model described in example ($1$) of the main text.
Writing $\ham_0 = \sum_{\bk} \hh_{\bk}$, the velocity operators are defined by
$\hv_{\nu} \equiv \sum_{\bk} \ptl \hh_{\bk}/\ptl k_{\nu}$ with $\nu = (x, y)$.
This leads to
\beq
\hGa_{A_{2g}} = \frac{2(t+t') V_0}{\bar{\om}} \sum_{\bk}   q_{\bk}
\left(\cda_{\bk, x} c_{\bk, y} - {\rm h.c.} \right),
\eeq
$q_{\bk} \equiv -\ep_m/V_0 \sin k_x \sin k_y (\cos k_x + \cos k_y) + 2 \sin^2 k_x \cos k_y + 2\cos k_x \sin^2 k_y$.
Likewise, computing
$\ptl^2  \hh_{\bk}/\ptl k_x^2 - \ptl^2  \hh_{\bk}/\ptl k_y^2$ we get
\begin{align}
\hGa_{B_{1g}} &=  2 \sum_{\bk} \big[ (t\cos k_x +t' \cos k_y) \cda_{\bk, x} c_{\bk, x} \nonumber \\
&- (t \cos k_y +t' \cos k_x) \cda_{\bk, y} c_{\bk, y} \big].
\end{align}

The next step is to write the above operators in the band basis for which we need the eigensystem of
the matrix associated with $\hh_{\bk}$. The eigenvalues are
\beq
\ep_{\bk, a/b} = m_{\bk} \pm \sqrt{n_{\bk}^2 + V_{\bk}^2},
\eeq
where $m_{\bk} \equiv (\ep_{\bk, x} + \ep_{\bk, y})/2$,
$n_{\bk} \equiv (\ep_{\bk, x} - \ep_{\bk, y})/2$, $V_{\bk}$ is the hybridization, and $(a, b)$ are the band
indices. The associated
eigenvectors are
\beq
\begin{pmatrix}
u_{\bk, a} \\ v_{\bk, a}
\end{pmatrix}
= \frac{1}{L_{\bk, a}}
\begin{pmatrix}
1 \\ s_{\bk, a}
\end{pmatrix},
\quad
\begin{pmatrix}
u_{\bk, b} \\ v_{\bk, b}
\end{pmatrix}
= \frac{1}{L_{\bk, b}}
\begin{pmatrix}
1 \\ s_{\bk, b}
\end{pmatrix},
\eeq
respectively,
where $s_{\bk, a} = V_{\bk}/(\sqrt{n_{\bk}^2 + V_{\bk}^2} + n_{\bk})$,
$L_{\bk, a}^2 = 1 + s_{\bk, a}^2$,
$s_{\bk, b} = V_{\bk}/(n_{\bk} - \sqrt{n_{\bk}^2 + V_{\bk}^2})$, and
$L_{\bk, b}^2 = 1 + s_{\bk, b}^2$.
The antihermitian property of $\hGa_{A_{2g}}$ ensures that it is an interband operator with
\begin{align}
\hGa_{A_{2g}} &= \frac{2(t+t') V_0}{\bar{\om}} \sum_{\bk}   q_{\bk} (u_{\bk, a}v_{\bk, b} - v_{\bk, a} u_{\bk, b}) \nonumber \\
&\times \left(\cda_{\bk, a} c_{\bk, b}  - {\rm h.c.} \right),
\end{align}
Likewise, for the Hermitian operator $\hGa_{B_{1g}}$ we get
\begin{align}
&\hGa_{B_{1g}} = 2\sum_{\bk} \Big [ (t\cos k_x +t' \cos k_y) u_{\bk, a}u_{\bk, b} - \nonumber \\
&-(t\cos k_y +t' \cos k_x) v_{\bk, a} v_{\bk, b} \Big ] \left( \cda_{\bk, a} c_{\bk, b}  + {\rm h.c.} \right) + \cdots,
\end{align}
where the ellipsis imply intraband operators that do not enter the calculation.
The evaluation of the cross-susceptibility follows standard steps, and we get
\begin{align}
\chi_{A_{2g}}^{B_{1g}}(i\Om_n) &= -\frac{2 (t+t')^2V_0}{\bar{\om}}
\frac{1}{\Om_0 \be} \sum_{\bk, \om_n} g_{\bk}
\nonumber \\
&\times
\left\{ G_b(\bk, i\om_n + i\Om_n) G_a(\bk, i\om_n)
- a \leftrightarrow b \right\}.
\end{align}
In the above
$g_{\bk} =  (\cos k_x + \cos k_y) V_{\bk} q_{\bk}/p_{\bk}$,
with $p_{\bk} = ((\ep_{\bk, x} - \ep_{\bk, y})^2/4 + V_{\bk}^2)^{1/2}$,
$G_{a/b}(\bk, i\om_n)  = 1/(i\om_n - \ep_{\bk, a/b})$. From the above we also have
$\chi_{A_{2g}}^{B_{1g}}(-i\Om_n) = -\chi_{A_{2g}}^{B_{1g}}(i\Om_n)$, which implies that
$\phi_{A_{2g}}^{B_{1g}}(i\Om_n) = 2\chi_{A_{2g}}^{B_{1g}}(i\Om_n)$.

\begin{figure}[t]
\begin{center}
\includegraphics[width=0.45\textwidth]{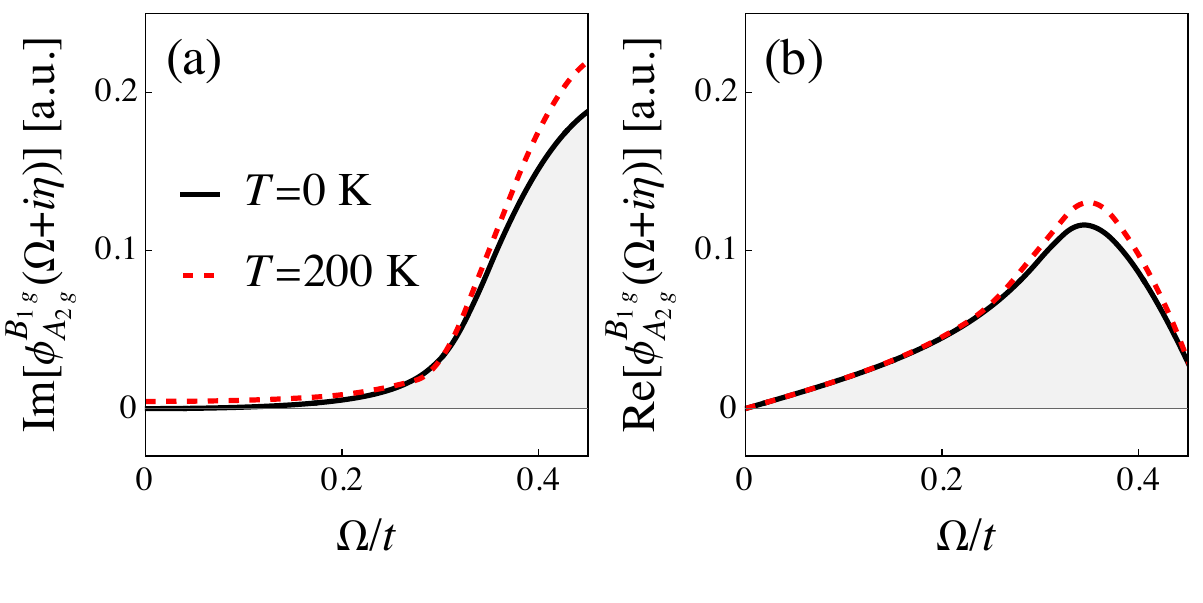}
\caption{(Color Online) Comparison between the imaginary (a) and real (b) part of
$\phi_{A_{2g}}^{B_{1g}}(\Om)$ at $T=0$ K (black line) and $T=200$ K (red dashed line). The low-temperature
limit is computed using Eq.\ (\ref{lowT}), the high temperature limit is obtained from Eq.\ (\ref{highT}) after
analytical continuation to real frequencies.}
\label{figS1}
\end{center}
\end{figure}

If we assume the electronic excitations to be undamped, the frequency sum can be performed analytically. We get
\begin{align}
L_{ab}(\bk, i\Om_n) &\equiv \frac{1}{\be} \sum_{\om_n} G_a(\bk, i\om_n) G_b(\bk, i\om_n + i\Om_n)
\nonumber \\
&= \frac{n_F(\ep_{\bk, a}) - n_F(\ep_{\bk, b})}{i\Om_n + \ep_{\bk, a} - \ep_{\bk, b}},
\label{highT}
\end{align}
where $n_F$ is the Fermi function. This gives,
\begin{align}
\phi_{A_{2g}}^{B_{1g}}(i\Om_n) &= -\frac{4 (t+t')^2V_0}{\bar{\om}}
\frac{1}{\Om_0} \sum_{\bk} g_{\bk} \left( n_F(\ep_{\bk, a}) - n_F(\ep_{\bk, b}) \right)
\nonumber \\
&\times
\left\{ \frac{1}{i\Om_n + \ep_{\bk, a} - \ep_{\bk, b}} + \frac{1}{i\Om_n + \ep_{\bk, b} - \ep_{\bk, a}} \right\}.
\nonumber
\end{align}
The other limit where the frequency sum can be performed analytically
is with finite electron lifetime but at $T=0$. In this limit
\begin{align}
L_{ab}(\bk, \Om + i\eta) &= \frac{1}{2\pi i}
\left[\frac{J_1(\bk, \om)}{\Om + \ep_{\bk, a} - \ep_{\bk, b}}
\right. \nonumber \\
&+  \left. \frac{J_2(\bk, \om)}{\Om +2i\ga + \ep_{\bk, a} - \ep_{\bk, b}} \right],
\label{lowT}
\end{align}
where
\begin{align}
J_1(\bk, \om) &= \ln(\Om - \ep_{\bk, b} + i \ga) + \ln(-\Om - \ep_{\bk, a} - i \ga)
\nonumber \\
&- \ln(- \ep_{\bk, a} + i \ga) - \ln(- \ep_{\bk, b} - i \ga),
\end{align}
\begin{align}
J_2(\bk, \om) &= -\ln(\Om - \ep_{\bk, b} + i \ga) - \ln(-\Om - \ep_{\bk, a} - i \ga)
\nonumber \\
&+ \ln(- \ep_{\bk, a} - i \ga) + \ln(- \ep_{\bk, b} + i \ga).
\end{align}
The computations for example ($1$) shown in the main text were performed using this limit.
Fig.\ \ref{figS1} shows that this is a good approximation for our model, since temperature is a
negligible energy scale compared to the Fermi energy.

For the numerical evaluation we set $t= 2$ eV, $t'=0.37$ eV, $\mu= 1.5$ eV, $V_0 = 0.32$ eV
and $\epsilon_m=0.35$ eV. The resulting Fermi surface is shown in Fig.\ 2(a) of the main text.
The threshold for interband transitions is given by $\Omega_{ph}=2V_{0}$
in the $\gamma\simeq 0$ limit, while it develops a tail at lower frequencies when the
electron broadening is increased. In Fig.\ 1(c)-(d) of the main text and Fig.\ \ref{figS1} we set $\gamma=0.05$ eV.

\subsection{Example ($2$), amplitude mode of the CDW phase}

Starting from the two-orbital model discussed in example ($1$), the CDW
phase can be effectively described by adding the mean-field inter-orbital interaction
$\mathcal{H}_{int}= \Delta \sum_{\mathbf{k}}[ c_{\mathbf{k+Q},x}^{\dag} c_{\mathbf{k-Q},y}
+ c_{\mathbf{k+Q},y}^{\dag} c_{\mathbf{k-Q}, x} + \text{h.c.}]$
to the total Hamiltonian $\mathcal{H}=\mathcal{H}_0 + \mathcal{H}_{int}$.
The amplitude mode is associated with amplitude fluctuations of the CDW order
parameter $\Delta$ around its equilibrium position. Its contribution to the Raman
response is obtained as the vertex correction in the amplitude channel of the bare
electronic response, which corresponds, from a diagrammatic perspective,
to the two-loop contribution shown in Fig.\ 1(b) of the main text.
The two fermionic susceptibilities in Eq.\~(8) of the main text explicitly read
\begin{equation}
\begin{aligned}
\chi_{A_{2g}}^{\Delta}(i\Omega_n) &= \frac{1}{\Omega_0 \beta}
\sum_{\mathbf{k}, \omega_n} \text{Tr} \left[\hat \Gamma_{A_{2g}}
\hat G(\mathbf{k},i\omega_n+i\Omega_n)\hat \Delta
\hat G(\mathbf{k},  i\omega_n) \right],\\
\chi_{\Delta}^{i}(i\Omega_n) &= \frac{1}{\Omega_0 \beta}
\sum_{\mathbf{k}, \omega_n} \text{Tr} \left[\hat \Delta
\hat G(\mathbf{k}, i\omega_n+i\Omega_n)\hat \Gamma_{i} \hat G(\mathbf{k}, i\omega_n) \right],
\label{bubble}
\end{aligned}
\end{equation}
where $i=(B_{1g}, B_{2g})$, $\hat \Delta \equiv \partial \mathcal{\hat H}_{int}/ \partial \Delta$
is the scattering vertex due to the amplitude mode and $\hat G(\mathbf{k}, i\omega)=
(i\omega - \mathcal{\hat H}_{\mathbf{k}})^{-1}$ is the electronic Green's function,
with $\mathcal{H}_{\mathbf{k}}$ written in the extended orbital basis defined by the
$\Psi_{\mathbf{k}} = \left( c_{\mathbf{k-Q},x}, c_{\mathbf{k+Q},x}, c_{\mathbf{k-Q},y},
c_{\mathbf{k+Q},y}  \right)^T$ spinor, such that $\mathcal{H}=\sum_{\mathbf{k}}
\Psi_{\mathbf{k}}^{\dag} \mathcal{\hat H}_{\mathbf{k}} \Psi_{\mathbf{k}} $.
Following the same procedure detailed in the previous Section, it is convenient to write
$\hat G$ in the band basis and perform the frequency sum at $T=0$. One then introduces
a finite electron lifetime $\gamma$, set to $\gamma=0.1\Delta$ with $\Delta=0.1$ eV in
Fig.\ 2 of the main text. The tight-binding parameters are the same detailed in example ($1$),
while the ordering wave-vector is set to $\mathbf{Q}=\sqrt{2} Q_0\{\cos[(1+r)\pi/4],
\sin[(1+r)\pi/4]\}$, with $Q_0=5\pi/8$. The amplitude mode propagator is well described
by a Lorentzian function $D(\Omega)= 2E_0 / [(\Omega+i\gamma_0)^2-E_0^2]$,
with $E_0=0.005$ eV  and $\gamma_0=0.2E_0$.

\end{document}